\documentclass[twocolumn,twoside,slac_two]{revtex4}
\usepackage{graphicx}
\usepackage{fancyhdr}
\begin{document}

\title{Wide-Field MAXI: soft X-ray transient monitor}


\author{Makoto Arimoto, Nobuyuki Kawai, Yoichi Yatsu}
\affiliation{Tokyo Institute of Technology, 2-12-1 Ookayama, Meguro-ku, Tokyo, 152-8551, Japan}
\author{Hiroshi Tomida, Shiro Ueno, Masashi Kimura}
\affiliation{Japan Aerospace Exploration Agency, 2-1-1 Sengen, Tsukuba, Ibaraki, 305-8505, Japan}

\author{Tatehiro Mihara, Motoko Serino, Mikio Morii}
\affiliation{RIKEN, 2-1 Hirosawa, Wako, Saitama, 351-0198, Japan}

\author{Hiroshi Tsunemi}
\affiliation{Osaka University, 1-1 Machikaneyama, Toyonaka, Osaka, 560-0043, Japan}

\author{Atsumasa Yoshida, Takanori Sakamoto}
\affiliation{Aoyama Gakuin University, 5-10-1 Fuchinobe, Chuo-ku, Sagamihara, Kanagawa, 252-5258, Japan}

\author{Takayoshi Kohmura}
\affiliation{Tokyo University of Science, 2641, Yamazaki, Noda, Chiba, 278-8510, Japan}

\author{Hitoshi Negoro}
\affiliation{Nihon University, 1-8-14 Surugadai, Chiyoda, Tokyo, 101-8308, Japan}

\author{Yoshihiro Ueda}
\affiliation{Kyoto University, Kitashirakawa-Oiwake-cho, Sakyo-ku, Kyoto, 606-8502, Japan}

\author{Yohko Tsuboi}
\affiliation{Chuo University, Kasuga 1-13-27, Bunkyo-ku,  Tokyo, 112-8551, Japan}

\author{Ken Ebisawa}
\affiliation{Institute of Space and Astronautical Science (ISAS), 3-1-1 Yoshinodai, Sagamihara, Kanagawa,  229-8510, Japan }

\begin{abstract}
{\it Wide-Field MAXI} ({\it WF-MAXI}: Wide-Field Monitor of All-sky X-ray Image) is a proposed mission to detect and 
localize X-ray transients including electro-magnetic counterparts of gravitational-wave events such as gamma-ray bursts and supernovae etc., which are expected to be directly detected for the first time in late 2010's by the next generation gravitational telescopes such as Advanced LIGO and KAGRA. The most distinguishing characteristics of 
{\it WF-MAXI} are a wide energy range from 0.7 keV to 1 MeV and a large field of view ($\sim$ 25 \% of the entire sky), which are realized by two main instruments: (i) Soft X-ray Large Solid Angle Camera (SLC) which consists of four pairs of crisscross coded aperture cameras using CCDs as one-dimensional fast-readout detectors covering 0.7 $-$ 12 keV and (ii) Hard X-ray Monitor (HXM) which is  a multi-channel array of crystal scintillators coupled with avalanche photo-diodes covering 20 keV $-$ 1 MeV.

\end{abstract}

\maketitle

\thispagestyle{fancy}


\section{Scientific goals}
{\it Wide-Field MAXI} ({\it WF-MAXI}: Wide-Field Monitor of All-sky X-ray Image) \cite{2014SPIE.9144E..2PK} on the ISS is a mission to detect and 
localize X-ray transients with a large field of view (FoV $\sim$25\% of the entire sky) 
covering a wide energy band from 20 keV to 1MeV, monitoring the entire sky.
The characteristic feature is a strong capability of detecting soft X-ray photons ($<$ 10 keV) from X-ray transients 
with a fine localization accuracy of $\sim$0.1$^\circ$, with a cadence of 90 min.
The transient search below 10 keV with the large FoV has been done with only a few satellites 
(e.g., HETE-2 \cite{2003AIPC..662....3R}, MAXI\cite{2009PASJ...61..999M}), 
so there is huge room for discovery space on  the high energy astronomy.

The most challenging target object of {\it WF-MAXI} is X-ray transients including electro-magnetic counterparts of
 gravitational-wave (GW) events 
such as gamma-ray bursts (GRBs) and supernovae (e.g., core-collapse SNe) which are expected to be directly detected for the first time 
in late 2010's by the next generation GW telescopes such as Advanced LIGO, Virgo and KAGRA.
%If, luckily, a core-collapse supernova (cc-SN) occurs in the Milky Way galaxy, our X-ray observation would contribute to understanding the anisotropic explosion mechanism of cc-SNe by help of observed neutrinos, in addition to gravitational wave.
However, 
the localization by the GW telescopes is too coarse ($\sim$10$^\circ$) to associate the detected GW sources with known 
astronomical objects, and/or measure their distances, and identify their physical origins. 
Soft X-ray band gives us  a promising channel considering the huge energy density at the source, 
and yet all-sky monitoring with sufficient sensitivity and cadence has never been performed.
If a GW event is detected by {\it WF-MAXI}, its localization will be performed with an positional accuracy of 0.1$^\circ$. After that, {\it WF-MAXI} issues its alert to the international astronomical community, 
which leads to enabling follow-up observations with  X-ray, optical and infrared observatories (e.g., ASTRO-H, Subaru, TMT, JWST etc.)  to measure its distance and study on its environment and progenitor.

A part of GW events is thought to  originate from compact-binary coalescence sources including neutron stars, stellar-mass black holes and intermediate-mass black holes.
Although there is a large uncertainty of expected GW event rate \cite{2010CQGra..27q3001A}, we show a summary of  
expected detection rates of GW events by current X-ray observatories  with a large FoV in Table $\ref{event_rate}$,
 assuming that 10 GW events happen in a year. 
 {\it WF-MAXI} has the highest detectability of GW sources among the current observatories.

\begin{table*}[!t]
\begin{center}
\caption{Expected detection rates of GW source by current X-ray observatories assuming that 10 GW events happen in a year}
\begin{tabular}{|l|c|c|c|c|}
\hline \textbf{Mission} & \textbf{FoV} [\%] & \textbf{operation ratio} & \textbf{expected detection rate} & \textbf{soft X-ray sensitivety}\\
 & ratio to 4$\pi str$ & [\%] & \textbf{of GW events} [events/year] & (below 10 keV)\\
\hline Swift/BAT & 11 & 80  & 0.88 & N/A \\
\hline MAXI & 2 & 40  & 0.08 & $\circ$ \\
\hline Integral IBIS & 0.2 & 100  & 0.02 & N/A \\
\hline \textbf{WF-MAXI} & 25 & 70  & \textbf{1.67} & \textbf{$\circ$} \\
\hline
\end{tabular}
\label{event_rate}
\end{center}
\end{table*}

Not only for GW events but also for energetic members of astrophysical objects, such as neutron star binaries, black hole binaries and active galactic nuclei 
(AGN), {\it WF-MAXI} detects the onset of  its activities and issues alerts to the astronomical community of the world (e.g., The Astronomer's Telegram).
Furthermore, {\it WF-MAXI} also detects short high-energy transients such as GRBs and tidal disruption events and short soft X-ray transients such as stellar flares, nova ignitions and supernova shock breakouts.

\section{Mission instruments}
\begin{figure*}[htbp]
\centering
\includegraphics[width=135mm]{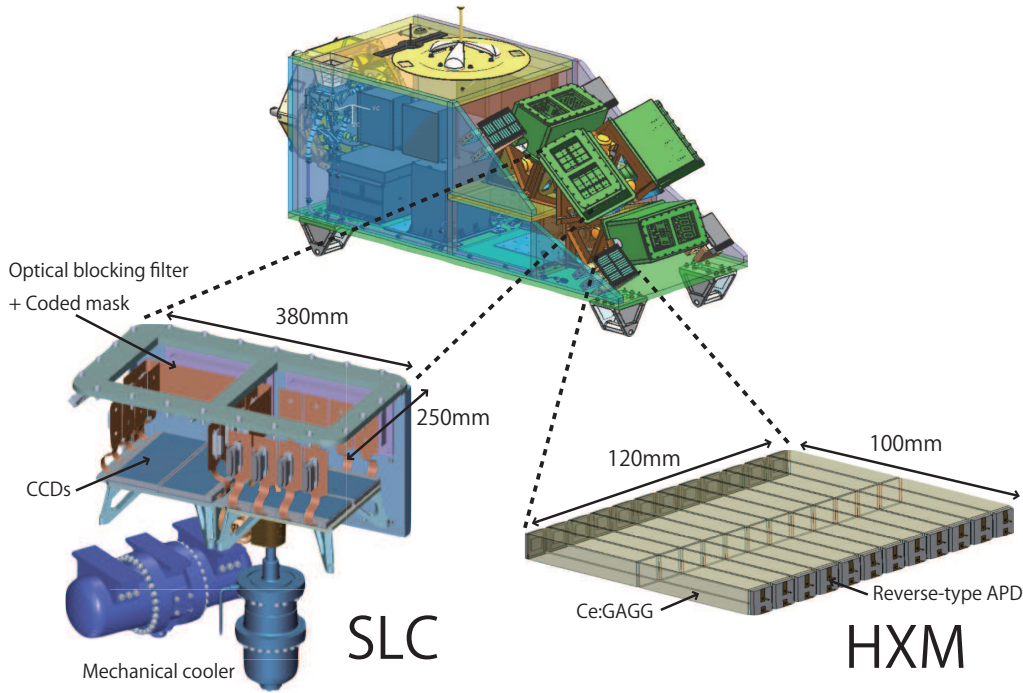}
\caption{Configuration of the {\it WF-MAXI} payload. Four modules of SLC and HXM are implemented at different
 four angles to cover 25 \% of the entire sky in the energy band from 0.7 keV to 1 MeV. SLC: Soft X-ray Large Solid
  Angle Camera, HXM: Hard X-ray Monitor} \label{WF-MAXI_overview}
\end{figure*}

{\it WF-MAXI} has two main instruments of Soft X-ray Large Solid Angle Camera (SLC) and Hard X-ray Monitor (HXM)
 to detect X-ray photon in the wide energy range of 0.7 keV to 1 MeV.  Four modules of SLC and HXM are mounted on the payload at different four angles to cover $\sim$25\% of the entire sky as shown in Fig.  \ref{WF-MAXI_overview}.

SLC and HXM are sensitive for
the energy band of 0.7 -- 12 keV  and 20 keV -- 1 MeV, respectively.
Both two instruments
share the same FoV. SLC plays an important role in localizing X-ray transients with an
accuracy of  $\sim$0.1$^\circ$. Furthermore for HXM it is quite important to observe GRBs with a wide X-ray band: GRBs' 
spectra are well represented by two powerlaw functions connected smoothly, which is called the Band function \cite{1993ApJ...413..281B}, and a
maximum peak energy $E_{\rm peak}$ in the $\nu F_\nu$ space is one of the fundamental quantities for GRBs. 
As the $E_{\rm peak}$ ranges from a few keV to a few MeV, HXM plays a crucial role in determining $E_{\rm peak}$ in the energy band from 20 keV to 1MeV. Combined with SLC, even $E_{\rm peak}$ of soft-class GRBs called “X-ray flashes” can be determined in the range down to a few keV. 
%The overview of the {\it WF-MAXI} instruments is shown in Fig. \ref{WF-MAXI_overview}.

\subsection{Soft X-ray Large Solid Angle Camera}
The primary scientific instrument of WF-MAXI is SLC \cite{2014SPIE.9144E..60K}, which has a capability of  detecting and localizing various soft X-ray ($<$ 10 keV) transients including possible GW counterparts, GRBs, SN shock breakouts, tidal disruption events, nova ignition, X-ray bursts, AGN flares, and stellar flares. In the energy
band  numerous characteristic X-ray lines (e.g. Ne, Mg, Si, S, Fe) exist to trace the environment of the progenitor or 
burst mechanism and these can be resolved by the energy
resolution of a CCD instrument. We therefore adopt a CCD as a position sensitive detector.
Coded mask is adopted for the localization, as it can achieve a large field of view without much technical difficulty. 

Since the imaging field of a CCD camera fixed to the ISS platform moves in the sky at an angular velocity
 of $\sim$0.1$^\circ$/s, we need to read out the image data on a timescale shorter than 1 $s$ (e.g., 0.1 $s$) to achieve $\sim$0.1$^\circ$ position accuracy. We therefore use one dimensional image from CCD
 with a time resolution of 0.1 $s$  for fast readout.
  We assign X and Y coordinates to a CCD plane where CCDs are vertically-aligned in two directions.
Thus, each module of  SLC  contains two arrays of CCD in X and Y directions, a pair of coded masks, 
  a part of the electronics that drives and reads out CCD's image data, a mechanical cooler and the chassis as shown in Fig.  \ref{WF-MAXI_overview}. 
  The dimensions of the camera module are 380mm $\times$ 250mm $\times$ 220mm without the mechanical cooler. 

We use 16 CCDs (Hamamatsu) for a SLC with an effective area of 293 cm$^2$ larger than that used in 
MAXI/SSC \cite{2011PASJ...63..397T}. 
The CCD is a similar model developed for ASTRO-H/SXI \cite{2014SPIE.9144E..29H} with some minor changes 
that include pixel format, PGA packaging (instead of wire bonding),
an addition of fiducial mark used for alignment with the coded mask and a surface processing on the CCD.
The surface of the CCD is covered with 150 $\sim$ 200 nm aluminum to block the optical light from optical sources and
 scattered lights from bright objects. 
 Both sides are coated with black colorant to prevent the infrared light leaking into the silicon CCD chip.
 Furthermore, we dispose a thin aluminum-coated polyimide layer at the camera window to 
 block  incoming heat and reflected sun light and He II ultraviolet emission from the upper atmosphere.

\begin{figure}[htbp]
\centering
\includegraphics[width=60mm]{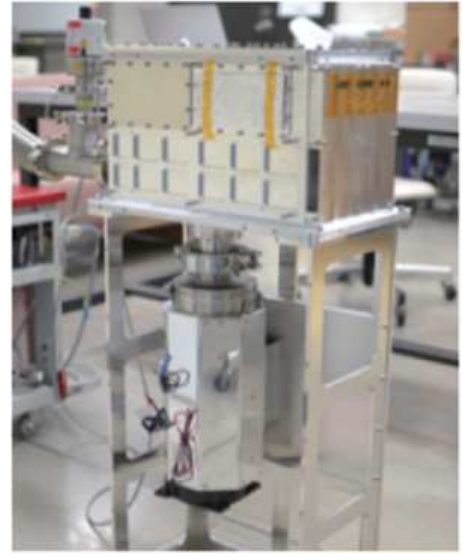}
\caption{Prototype model of the mechanical cooler for SLC} \label{cooler}
\end{figure}

Cooling 16 CCD chips to −100$^\circ$C on the ISS payload is a critical task for our mission to assure the CCD performance.
 As {\it WF-MAXI} has no attitude  control system, the payload will be illuminated by the sun light every orbit ($\sim$90 minutes). 
  There is no place for radiators permanently facing the deep space to release the heat. Then development of a thermal model for the SLC module is in progress and verification of its feasibility was almost achieved.
We find that the dominant heat paths to the CCD contribute from conductances through flexible cables to the CCD packages, the support legs of the base plate,  cold plate to bus interface plate and the radiation from the flexible cables, while the heat production on the CCD itself is small. Taking account these heat paths, the target temperature of the CCD is achievable. However the four mechanical coolers consume a significant amount of power ($>$ 300 W). Further design study of the conductances which attribute to the critical thermal path such as  
CCD flexible cables (e.g., use of thinner conductive wires) or relaxation of the required temperature by improving the CCD dark current (e.g.,  surface processing on the CCDs), is underway. Especially, the required heat load for the mechanical cooler is estimated to be 5.2 W and we plan to verify the thermal model with a prototype model by 2015 (Fig. \ref{cooler}). 

\subsection{Hard X-ray Monitor}

As the secondary scientific instrument of {\it WF-MAXI}, HXM \cite{2014SPIE.9144E..5ZA} measures the energy spectra and light curves of short transient events in the 20 keV – 1 MeV energy range and provides the trigger for GRBs.

HXM consists of a 24-channel array of 
Ce-doped Gd$_3$Al$_2$Ga$_3$O$_{12}$ (Ce:GAGG) scintillator coupled with avalanche
 photodiode (APD) covering the hard X-ray band with an effective area of  120 cm$^2$ (Fig. \ref{WF-MAXI_overview}). 
To obtain a better signal to noise (S/N) ratio and detect higher-energy photons, 
we select the Ce:GAGG crystal due to its high light yield 
(46,000 photons/MeV) and density (6.63 g/cm$^3$), where scintillation light peaks at a wavelength of 520 nm, in well matching with the sensitivity of the silicon photon detector. The lower energy threshold of 20 keV is achievable by operating it at −20 $\sim$ 0 $^\circ$C using a passive thermal structure or a thermoelectric cooler.

We adopt flight-proven reverse-type APDs with a pixel size of 5 $\times$ 5 mm$^2$ provided by Hamamatsu Photonics to detect scintillation lights of the Ce:GAGG crystal.
The performance of the APD is low-noise and flight-proven to be radiation hardness  on CubeSat ({\it Cute-1.7+APD II} \cite{2010JGRA..115.5204K} ) working in a polar orbit for five years as a radiation particle monitor. 
Its technology is also adopted for micro-satellite  {\it Tsubame} \cite{2014SPIE.9144E..0LY}  and ASTRO-H \cite{2014SPIE.9144E..27S}.
In addition, as it is well known that the gain of  APDs strongly depends on temperature and the bias voltage, in the HXM system the APD gain dependent on temperature  is controlled by adjusting the bias voltage.

\begin{figure}[htbp]
\centering
\includegraphics[width=60mm]{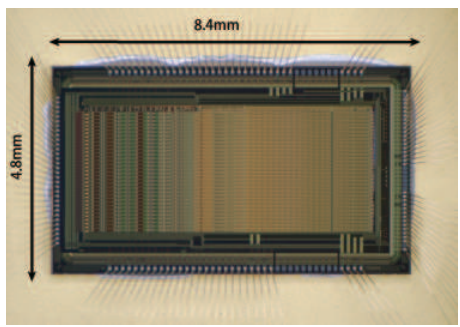}
\caption{Developed LSI dedicated for processing  APD signals (HXM). The LSI contains 32-ch analog amplifiers and AD converters and the design of the noise suppression is implemented. } \label{TT01APD32}
\end{figure}

\begin{figure}[htbp]
\centering
\includegraphics[width=60mm]{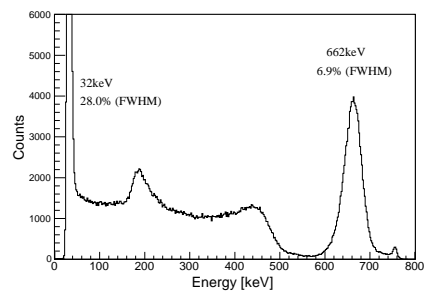}
\caption{Energy spectrum of $^{137}$Cs with the reverse-type APD (S8664-55) coupled to the Ce:GAGG crystal scintillator. } \label{Cs137spec}
\end{figure}

We developed a new LSI dedicated for an analog amplification of APDs' signal.
The new LSI contains 32-channel amplifiers and AD converters with a chip size of 4.8 $\times$ 8.4 mm$^2$ (Fig. \ref{TT01APD32}).
Especially, to accomplish the quick development we utilized  well-studied 0.35 $\mu$m CMOS technology based on Open IP project by Professor Hirokazu Ikeda and
accumulated knowledge for a decade. 
As a detector capacitance of APDs
is large ($\sim$100 pF), its capacitive noise  is crucial for detection of X-ray photons around the lower energy threshold ($\sim$20 keV).
We thus designed the analog circuit to suppress the capacitive noise (e.g., larger transconductance, larger gate area of input FETs and so on).
We show the performance  of the developed LSI in Table \ref{table_LSI} and one of obtained spectra in Fig \ref{Cs137spec}. 
Signals from 32 keV and 662 X-rays are clearly detected and its energy resolutions (FWHM) are determined to be 28.0\% and 6.9  \%, respectively.  The detection of 32 keV X-rays shows the low-noise amplifier in the new LSI almost has achieved the lower energy threshold of 20 keV in HXM.

\begin{table}[!t]
\begin{center}
\caption{Specification \& performance of the new LSI for HXM}
\begin{tabular}{lcc} \hline\hline\\[-6pt]
Number of channels        & 32  \\   
Dynamic range &  0 -- 300 fC\\
Non linearity & $<$4\% \\
Peaking time for trigger & 0.5 $\mu$s\\
Peaking time for spectroscopy & 3 $\mu$s\\
Equivalent Noise Charge & $\sim$2400 $e^-$ \\
Power supply & $\pm$1.65 V  \\
Power consumption    & $\sim$100 mW   \\   \hline 
\end{tabular}
\label{table_LSI}
\end{center}
\end{table}

\section{Summary}
{\it WF-MAXI} is a proposed mission of X-ray transient monitor  as a payload on the ISS. Its science goal is to detect and localize  X-ray transient sources and issue prompt alerts to the astrophysical community all over the world. The X-ray counterpart of the first directly detected GW event is the prime target of the {\it WF-MAXI}
 mission. Furthermore, it is the first dedicated transient monitor mission that covers a significant fraction ($\sim$25 \%) of the entire sky in the soft X-ray band with a energy resolution of CCD plus the hard X-ray band, which promises to open a new discovery space. 
  
  We have been developing  two mission instruments of SLC – Soft X-ray Large Solid Angle Camera and HXM – Hard X-ray Monitor. For SLC, the thermal design of  cooling the CCD chips to -100$^\circ$C, the prototype and its readout electronics are being developed. For HXM, the new LSI dedicated for the readout of signals from APDs was developed and we find that the designed low-noise analog amplifier achieved our goal of the lower energy threshold (20 keV).
  We will apply the {\it WF-MAXI} mission or modified mission to  Small-size project 2015 funded from JAXA to  
  develop and launch the payload for the beginning of the operation of next generation gravitational-wave observatories.

\bigskip % extra skip inserted
\begin{acknowledgments}
This work is supported by Grant-in-Aid for Scientific Research on Innovative Areas “New development in astrophysics through multi-messenger observations of gravitational wave sources”, MEXT KAKENHI Grant Number 80195031.
\end{acknowledgments}

\bigskip % extra skip inserted

\end{document}